\documentclass[preprint]{emulateapj}
\newcommand{\gsim}{\mbox{$\stackrel {>}{_{\sim}}$}} 
\newcommand{\lsim}{\mbox{$\stackrel {<}{_{\sim}}$}} 
 
\slugcomment{Accepted to the Astronomical Journal}
 
\shorttitle{Cyanoacetylene in IC 342}
\shortauthors{Meier, Turner \& Schinnerer} 
\begin{document} 
 
\title{Cyanoacetylene in IC 342: An Evolving Dense Gas Component with Starburst Age\footnote{\scriptsize Based on observations carried out with the IRAM Plateau de Bure Interferometer. IRAM is supported by INSU/CNRS (France), MPG (Germany) and IGN (Spain).}}

\author{David S. Meier\altaffilmark{1,2}, Jean L. Turner\altaffilmark{3} and 
Eva Schinnerer\altaffilmark{4}}

\altaffiltext{1}{New Mexico Institute of Mining and Technology, 802 Leroy Place, 
Socorro, NM 87801; dmeier@nmt.edu}
\altaffiltext{2}{Adjunct Assistant Astronomer, National Radio Astronomy Observatory,
P. O. Box O, 1003 Lopezville Road, Socorro, NM 87801}
\altaffiltext{3}{Department of Physics and Astronomy, UCLA, Los Angeles, 
CA 90095--1562; turner@astro.ucla.edu}
\altaffiltext{4}{Max-Planck-Institut f\"{u}r Astronomie, K\"{o}nigstuhl 17, D-69117 Heidelberg, 
Germany; schinner@mpia.de}

\begin{abstract} 
We present the first images of the J=5--4 and J=16--15 lines of the dense gas tracer, 
cyanoacetylene, HC$_{3}$N, in an external galaxy. The central 200 pc of the nearby 
star-forming spiral galaxy, IC~342, was mapped using the VLA and the Plateau de
Bure Interferometer.  HC$_{3}$N(5--4) line emission is found across the nuclear
mini-spiral, but is very weak towards the starburst site, the location of the strongest 
mid-IR and radio emission.  The J=16--15 and 10--9 lines are also faint near the large 
HII region complex, but are brighter relative to the 5--4  line, consistent with higher excitation. 
The brightest HC$_{3}$N emission is located in the northern arm of the nuclear minispiral,
100 pc away from the radio/IR source to the southwest of the nucleus. This
location appears less affected by ultraviolet radiation, and may represent 
a more embedded, earlier stage of star formation.  HC$_{3}$N excitation temperatures 
are consistent with those determined from C$^{18}$O; the gas is dense,  
$10^{4-5}~\rm cm^{-3},$ and cool, T$_{k}~< 40$ K. So as to not violate limits on 
the total H$_{2}$ mass determined from C$^{18}$O, at least two dense components 
are required to model IC 342's giant molecular clouds.  These observations suggest that 
HC$_{3}$N(5--4) is an excellent probe of the dense, quiescent gas in galaxies. 
The high excitation combined with faint emission towards the dense molecular 
gas at the starburst indicates that it currently lacks large masses of very dense gas.  
We propose a scenario where the starburst is being caught in the act of dispersing or 
destroying its dense gas in the presence of the large HII region.  This explains  
the high star formation efficiency seen in the dense component.  The little remaining 
dense gas appears to be in pressure equilibrium with the starburst HII region.
\end{abstract} 
\keywords{galaxies:individual(IC 342) --- galaxies:starburst ---
  galaxies: ISM --- radio lines}
 
\section{Introduction \label{intro}} 

Little is known about the properties of the dense gas component on GMC 
scales in external galaxies yet it is dense gas
($n_{H_{2}}> 10^{4}$ cm$^{-3}$) that most directly correlates with
global star formation rates \citep[e.g.,][]{GS04,WEGSSV05}.  To
understand the evolution and regulation of current star formation in
galaxies, dense gas properties need to be accurately characterized.

In a recent imaging survey of millimeter-wave molecular lines in the nearby
spiral nucleus of IC 342, a surprising degree of morphological variation
is observed in its dense gas \citep[][]{MT05}, which was not anticipated from 
CO or HCN \citep[e.g.,][]{I90,DRGGGM92}. These and other
observations \citep[e.g.,][]{GMFN00,GMFN01,GMFUN02,UGFMR04,UGMFN06}
confirm that CO and HCN fail to fully constrain excitation and chemical 
properties of the dense clouds.  Various probes 
of the dense component exist, each with various strengths and weaknesses
\citep[e.g.,][]{MT05,P07,Bet08,NCSDHW08,KNGMCGE08,GGPFU08}.
HCN(1--0) is the most commonly used dense gas tracer, primarily
because it is bright, however, it is optically thick in most starburst
environments \citep[e.g.][]{DRGGGM92,MT04, KWWBRM07}, has a widely spaced 
rotational ladder with its J$>$3 transitions in the submm and can be sensitive
to chemical effects \citep[PDR effects and IR pumping;
e.g.][]{FMCB93,APHC02}.

Here we present a study of the physical conditions of the very dense
gas component of a nearby starburst nucleus using HC$_{3}$N.  
With its large electric dipole moment ($\mu$ = 3.72 debye
versus 3.0 debye for HCN), low opacity and closely spaced rotational
ladder accessible to powerful centimeter and millimeter-wave
interferometers, HC$_{3}$N is well-suited to high resolution imaging
of the structure and excitation of the densest component of the molecular gas 
\citep[eg.][]{MTPZ76,VLSW83,AMMMB11}. HC$_3$N(5--4) has a critical density of $5\times 
10^{4}$~cm$^{-3}$ and an upper level energy of E/k = 6.6 K; the 16--15
line has a critical density of $5\times 10^{5}$~cm$^{-3}$ and an upper
level energy of E/k = 59 K.

We have observed the J = 5--4 and J = 16--15 lines of HC$_{3}$N at 
$<$2\arcsec\ resolution in IC 342 with the Very Large Array\footnote{The 
National Radio Astronomy Observatory is a facility of the National Science Foundation 
operated under cooperative agreement with by Associated Universities, Inc.}
(VLA) and the Plateau de Bure Interferometer (PdB), respectively. IC
342 is one of the closest \citep[D$\sim$3 Mpc, or 2\arcsec\ = 29 pc;
eg.][]{SCH02,K05}, large spirals with active nuclear star formation
\citep[][]{BGMNSWW80,TH83}. It is nearly face-on, with bright
molecular line emission both in dense clouds and a diffuse medium
\citep[][]{Letal84,I90,DRGGGM92,TH92}. The wealth of data and
proximity of IC~342 permit the connection of cloud properties with
star formation at sub-GMC spatial scales. The data published here
represent the first high resolution maps of HC$_3$N(5--4) and HC$_3$N(16--15) 
in an external galaxy. Excitation is a key component
in the interpretation of molecular line intensities. So these two maps
are compared with the previously published lower resolution
HC$_{3}$N(10--9) map made with the Owens Valley Millimeter Array
\citep[][]{MT05}, to constrain the physical conditions of the densest
component of the ISM in the center of IC342, and to correlate properties 
of the dense gas with star formation, diffuse gas, and chemistry.

\section{Observations \label{obs}} 
 
Aperture synthesis observations of the HC$_{3}$N J=5--4 rotational
line (45.490316 GHz) towards IC 342 were made with the D configuration
of the VLA on 2005 November 25 (VLA ID: AM839). The synthesized beam is
1\farcs95$\times $1\farcs5 (FWHM); position angle (pa)=-11.2\degr. Fifteen~1.5625 MHz
channels were used, for a velocity resolution of 10.3 km s$^{-1}$,
centered at $v_{LSR}=30$~ km s$^{-1}$. The phase center is
$\alpha$(J2000) = 03$\rm ^{h}$46$\rm ^{m}$48\fs 3; $\delta$(J2000) =
68\degr 05\arcmin 47\farcs0. Amplitude tracking and pointing was done
by observing the quasar 0228+673 every 45 minutes. Phases were
tracked by fast switching between the source and the quasar 0304+655
with 130s/50s cycles. Absolute flux calibration was done using 3C48
and 3C147 and is good to $\sim$5--10\%. Calibration and analysis was
performed with the NRAO Astronomical Image Processing Software (AIPS)
package. The naturally weighted and CLEANed datacube has an rms of 0.70 
mJy beam$^{-1}$. Correction
for the primary beam ($\sim$1\arcmin\ at 45 GHz) has not been applied.
The shortest baselines in the dataset are $\simeq$5 k$\lambda$,
corresponding to scales of $\sim$40\arcsec; structures larger than
this are not well-sampled.

The HC$_{3}$N(16-15) line emission in the center of IC 342 was
observed with 5 antennas using the new 2 mm receivers of the IRAM
Plateau de Bure interferometer (PdBI) on 2007 December 28 and 31 in C
configuration with baselines ranging from 24 to 176\,m. The phase
center of the observations was set to $\alpha$(J2000) = 03$\rm
^{h}$46$\rm ^{m}$48\fs 105; $\delta$(J2000) = 68\degr 05\arcmin
47\farcs84. NRAO\,150 and 0212+735 served as phase calibrators and were
observed every 20 min. Flux calibrators were 3C84 and 3C454.3. The
calibration was done in the GILDAS package following 
standard procedures. The HC$_{3}$N(16-15) line at 145.560951\,GHz was
observed assuming a systemic velocity of $v_{LSR}=46$~ km s$^{-1}$ and
a spectral resolution of 2.5\,MHz (5.15 km s$^{-1}$). The average 2 mm continuum was
obtained by averaging line-free channels blue and redward of the
HC$_{3}$N and H$_{2}$CO lines and subtracted from the $uv$ datacube to
obtain a continuum-free datacube. The final naturally weighted and
CLEANed datacube with 10\,km\,s$^{-1}$ wide channels has a CLEAN beam
of 1.83''$\times$1.55'', pa=46$^o$ and an rms of 1.7 mJy/beam.

Single-dish observations of HC$_{3}$N(16--15) find a peak brightness temperature 
of 6.7 mK in a 16.9$^{''}$ beam \citep[][]{AMMMB11}.  Convolving our map to this 
beam and sampling it at the same location yields a peak brightness of 7.1 mK, 
which agrees within the uncertainties of both datasets.  Therefore no flux is 
resolved out of the interferometer maps, as expected for these high density 
tracers.

\section{Results}
\subsection{The Dense Cloud Morphology \label{morp}}

\begin{figure}
\epsscale{0.99}
\plotone{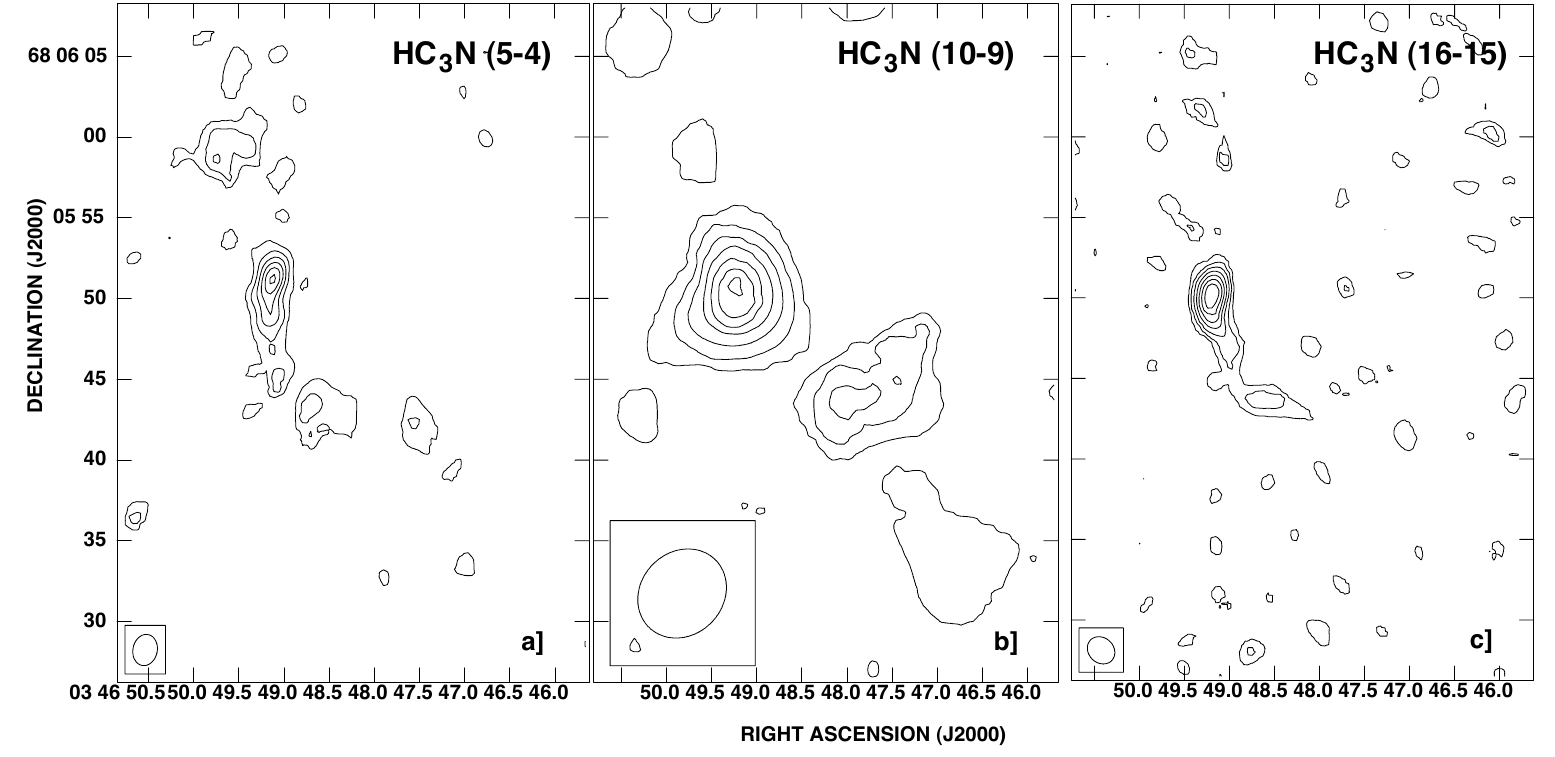}
\caption{The continuum subtracted HC$_{3}$N integrated intensity maps for IC 342. 
 {\it a)} The HC$_{3}$N(5--4) contoured in steps of 7.3 K km
 s$^{-1}$  (3$\sigma$) for a resolution of $1.^{''}95 \times 1.^{''}50$. {\it b)}
 a continuum subtracted version of HC$_{3}$N(10--9) from \citet[][]{MT05} contoured 
 in steps of 1.25 K km s$^{-1}$ for a beam size of $5.^{''}9\times5.^{''}1$. {\it
 c)} HC$_{3}$N(16--15) contoured in 3$\sigma$ steps of 5.1 K km
 s$^{-1}$ for a resolution of $1.^{''}83 \times 1.^{''}55$.
\label{inti} }
\end{figure}

\begin{figure}
\epsscale{0.93}
\plotone{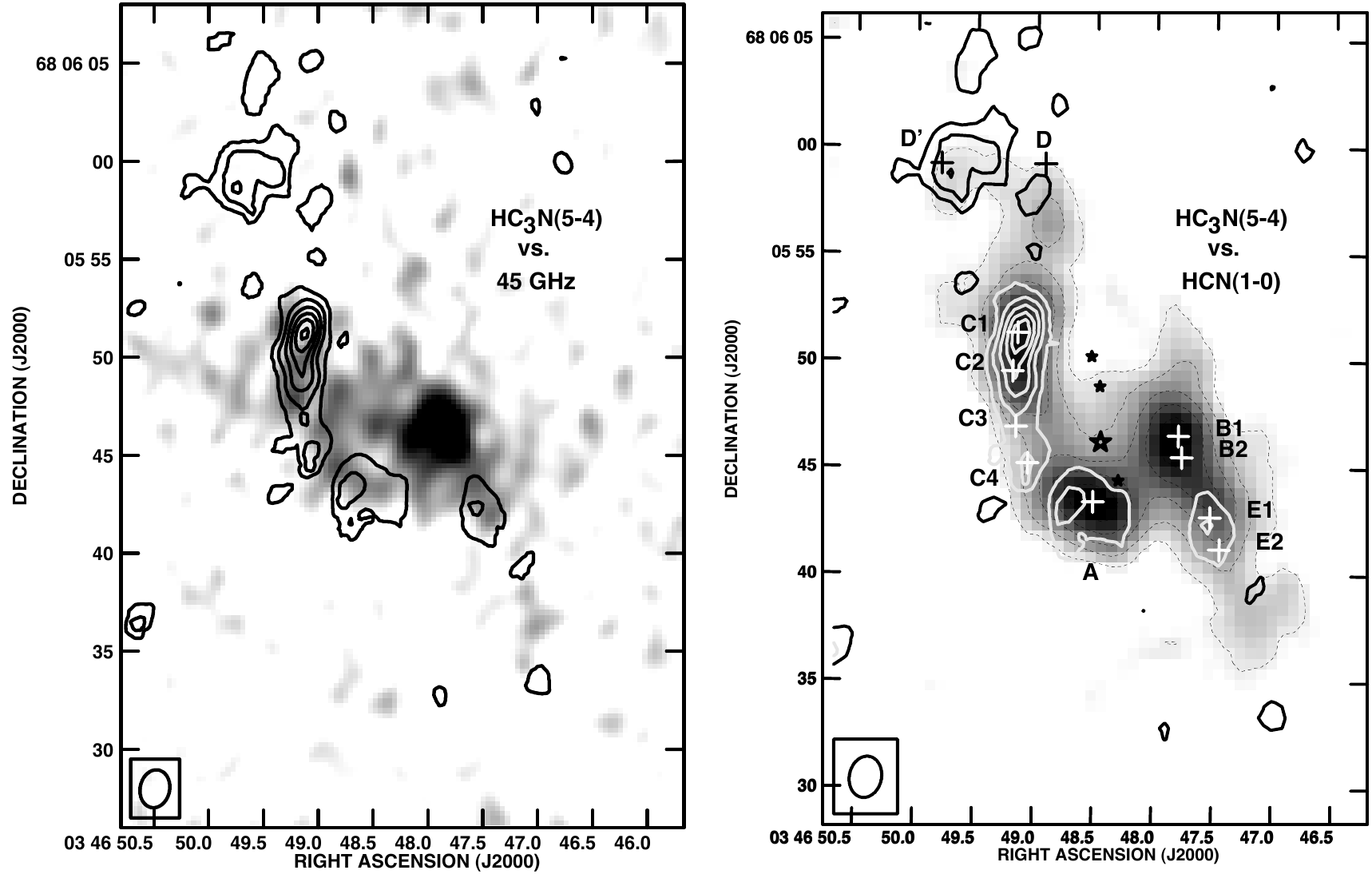}
\caption{{\it a)} The HC$_{3}$N(5--4) integrated intensity map {\it
    (black line)} for IC 342 overlaid on the 7 mm continuum emission
    {\it (grayscale)} extracted from the line free portion of the
    HC$_{3}$N(5--4) datacube. The continuum grayscale ranges from
    0.22 mJy bm$^{-1}$ - 2.2 mJy bm$^{-1}$.  HC$_{3}$N contours are as
    in Figure \ref{inti}. {\it b)} The HC$_{3}$(5--4) integrated intensity 
    map overlaid on the HCN(1--0) emission \citep[grayscale and dashed 
    contours; ][]{DRGGGM92}.  HC$_{3}$N contours are as in {\it a)}.  The 
    HCN(1--0) grayscale runs from 0.50 Jy beam km s$^{-1}$ to 4 Jy beam km s$^{-1}$ 
    and contours are in steps of 0.75 Jy beam km s$^{-1}$.  Positions of GMC cores \citep[][]{MT01} 
    and optical clusters \citep[e.g.][with absolute optical positions having absolute uncertainties of $
    \sim1^{''}$]{SBM03} are labelled.\label{inticont} }
\end{figure}

\begin{figure}
\epsscale{0.93}
\plotone{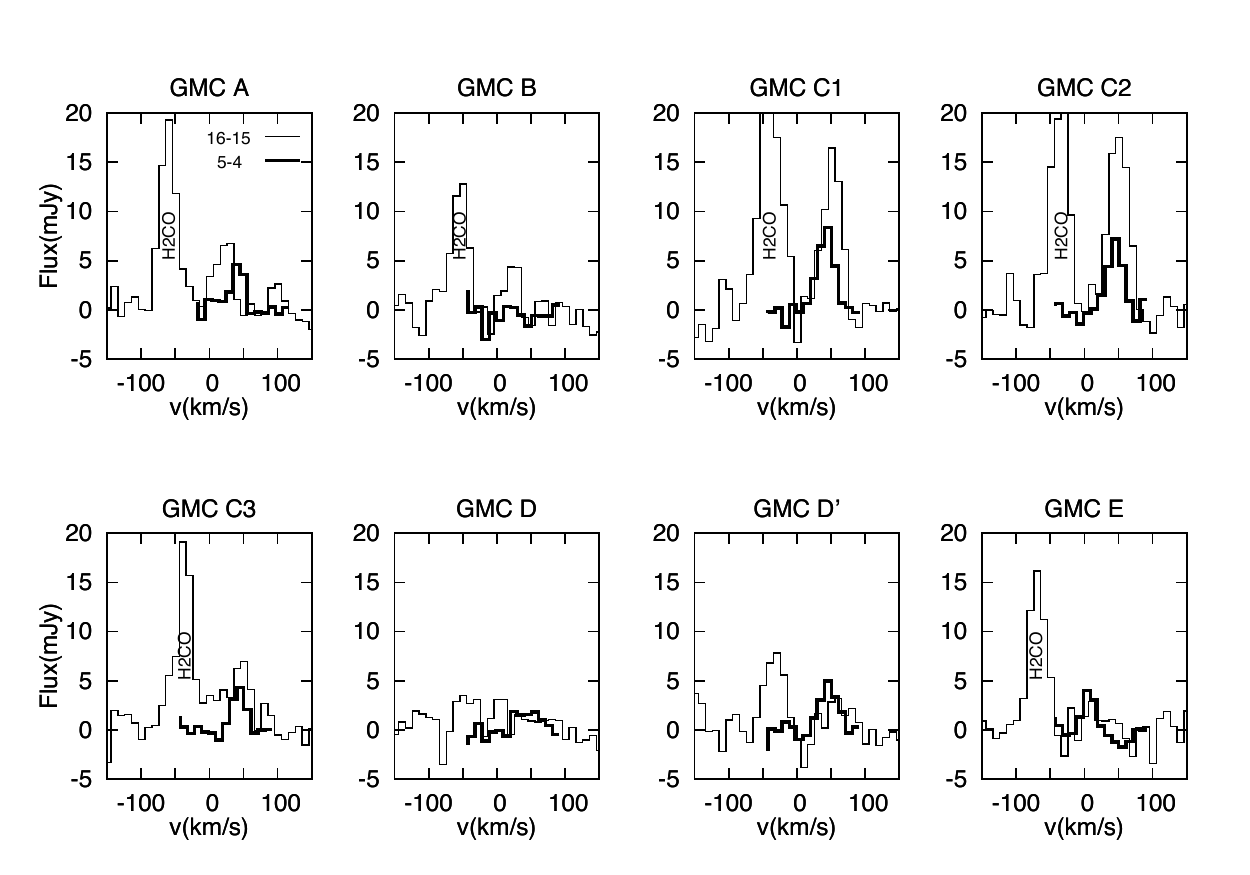}
\caption{The HC$_{3}$N(16--15) {\it (thin line)} and HC$_{3}$N(5--4)
{\it (thick line)} spectra. Spectra are summed over 2\arcsec\ apertures 
centered on each position at the locations of the main GMCs. Note that 
the bright spectral feature at $\sim$-60 km s$^{-1}$ is H$_{2}$CO($2_{02}-1_{01}$). 
\label{spec} }
\end{figure}

Continuum subtracted integrated intensity line maps of
HC$_{3}$N(5--4), HC$_{3}$N(10--9) and HC$_{3}$N(16--15) in IC~342 are
shown in Figure \ref{inti}. Figure \ref{inticont} shows the 5--4
transition overlaid on the 7~mm continuum image generated from offline
channels. Locations of giant molecular cloud (GMC) cores
\citep[][]{DRGGGM92,MT01} and the optical clusters \citep[e.g.][]{SBM03}, compared 
with HC$_{3}$N(5--4) and HCN(1--0) \citep[][]{DRGGGM92}, are also shown in
Figure \ref{inticont}. Figure \ref{spec} shows the HC$_3$N(5--4) and
HC$_{3}$N(16-15) spectra (in flux units) taken over the same 2\arcsec\
aperture centered on each cloud.

HC$_{3}$N(5--4) emission picks out most clouds seen in other dense gas
tracers (e.g HCN(1--0). The only labeled cloud not clearly detected in HC$_3$N(5--4)
is GMC~B, the cloud associated with the nuclear star-forming region
\citep[$\rm L_{IR}\sim 10^8~L_\odot$; ][]{BGMNSWW80,TH83}. Positions
of the GMCs measured in HC$_{3}$N are consistent with those fitted in
C$^{18}$O(2-1) to within a beam \citep{MT01}. An additional GMC is
detected in HC$_{3}$N(5--4) just south of GMC C3, labeled C4.  Unlike C$^{18}$O(2-1) which
peaks at GMC C2, HC$_{3}$N(5--4) emission peaks farther north, towards
GMC C1, at a distance of 105 pc from the nucleus suggesting changes in 
excitation across GMC C. GMCs A and D' are resolved into two components, 
but we do not discuss them as separate entities given the lower 
signal-to-noise, other than to state that there is a clear difference in velocity 
centroid between the 5--4 and 16--15 transitions (e.g. Figure \ref{spec}) 
indicating there is likely an excitation gradient across GMC A.  
No emission $>2\sigma$ is detected towards the weak CO(2-1) 
feature associated with the central nuclear star cluster \citep[][]{SBM03}. 

The J=16--15 line intensity is comparable to that of J=5--4 line in all clouds. 
While dominated by GMC C, HC$_{3}$N(16-15) is detected (or tentatively
detected) towards all clouds, except GMC D. Fluxes for GMCs D and D' 
are quite uncertain since they are just inside the half power point
of the PdBI primary beam.  That GMC C is much brighter than the other clouds 
in both transitions indicates large quantities of dense gas are present in this 
cloud.  HC$_{3}$N(16--15) favors C2 over C1. 
GMC B, while still faint in an absolute sense, is significantly brighter in
HC$_{3}$N(16--15) relative to 5--4. This is not unexpected given the
higher excitation requirements of the 16--15 line and that both B and
C2 have 7 mm continuum sources associated with massive star formation.

\subsection{Gas Excitation and the HC$_{3}$N $\Delta$ J Line Ratios  \label{ratio}}

Comparisons of the HC$_3$N (5--4), (10--9) and (16--15) maps
provide a chance to establish the excitation of the densest 
molecular cloud gas  component.  Line intensities from the J=5--4 and J=16-15 lines were
measured over 2\arcsec\ apertures centered on each of the GMCs. Table
\ref{Tinti} records gaussian fits to each spectrum along with
integrated line intensities. Peak antenna temperature ratios are
calculated for HC$_{3}$N(16--15)/HC$_{3}$N(5--4), hereafter denoted
R$_{16/5}$. R$_{16/5}$ ranges from less than 0.06 up to $\sim$0.5,
with GMC B having the highest value (Table \ref{Trat}). The 5--4 and
16--15 transitions generally bracket the peak of the level
populations, so we achieve good constraints on gas excitation. LTE
excitation temperatures, T$_{ex}$, implied by R$_{16/5}$ range from
$<$10 K to $>$18 K (where we have neglected T$_{cmb}$ in this
determination). Excitation is lowest towards GMCs D, D' and E. These
excitation temperatures T$_{ex}$ are similar to those found from the
presumably much less dense gas traced in C$^{18}$O \citep[Table
2;][]{MT01}. Only the GMC C clouds have significantly higher T$_{ex}$
in HC$_3$N --- towards C1 - C3 T$_{ex}$(HC$_{3}$N) are about a factor
of two greater than T$_{ex}$(C$^{18}$O).

The resolution of the HC$_{3}$N(10--9) data is significantly
lower than it is for the 5--4 and 16--15 lines. 
To compare J=5--4 and J=10-9 line intensities, the 
5--4 data were convolved to the resolution of the (10--9) data
\citep[$5.\arcsec9\times5.\arcsec1$][]{MT01}, then integrated
intensity ratios, hereafter R$_{10/5}$, were sampled at the locations
of R$_{16/5}$. Though at lower resolution than R$_{16/5}$, we make the
approximation that R$_{10/5}$ does not change on these sub-GMC scales.  
While leading to larger uncertainties, this provides a way to include all three transitions in
modeling dense gas excitation at very high
resolution. $R_{10/5}$ range from 0.25 to 1.4 (Table \ref{Tinti}).
GMC B has the highest value of 1.4. The remainder of the GMCs have
ratios of $0.3 < R_{10/5}< 0.6$. Excitation temperatures implied by
these ratios range from T$_{ex} = $ 6--16 K, consistent with those
derived from $R_{16/5}$ separately. T$_{ex}$ derived from
R$_{10/5}$ are similar to those derived from
C$^{18}$O. The only exception here is GMC A, the cloud where PDRs 
(Photon-Dominated Regions) dominate \citep[][]{MT05}. Towards GMC A
T$_{ex}$(C$^{18}$O) and T$_{ex}$(R$_{16/5}$) are 12 - 13 K, twice that of
T$_{ex}$(R$_{10/5}$).

Before modeling the densities implied by the line ratios, we test whether 
IR pumping can be responsible for the observed excitation.  IR pumping 
is important if, 
\begin{equation}
^{vib}B_{ul}~I_{\nu}(45 \mu m) ~\gsim~ ^{rot}A_{ul}
\label{pump}
\end{equation}
where $^{vib}B_{ul}$ is the Einstein $B_{ul}$ of the corresponding $\nu_{7}=1$ vibrational 
transitions at $\sim 45 ~\mu$m, $I_{\nu}$(45 $\mu$m) is the 45 $\mu$m intensity as 
seen by the HC$_{3}$N molecules and $^{rot}A_{ul}$ is the Einstein $A_{ul}$ of the 
rotational state \citep[e.g.][]{CA10}.  It is extremely difficult to estimate the applicable 
45 $\mu$m IR intensity, but it is expected to be most intense towards the starburst GMC (B).  
A detailed assessment of IR pumping must await high resolution MIR maps, but we 
constrain  $I_{\nu}$(45 $\mu m$) in several ways.  First, we take the 45 $\mu$m flux 
from \citet[][]{Brandl06} and scale it by the fraction of total 20 $\mu$m flux that comes from 
within 2.1$^{''}$ of the starburst as found by \citet[][]{BGMNSWW80} and then average over that 
aperture.  For the average $I_{\nu}$(45 $\mu$m) calculated this way,  
$^{vib}B_{ul}~I_{\nu}$(45 $\mu$m) is $10^{-4}$ ($10^{-2.5}$) times too low to pump the 
16--15 (5--4) transition.  Alternatively if we (very conservatively) take the total MIR luminosity 
from the central 30$^{''}$ and assume it comes from a blackbody of the 
observed color temperature \citep[e.g. $\sim$50 K;][]{BGMNSWW80} then $I_{\nu}$(45 $\mu$m) is 
still at least an order of magnitude too low to meet the inequality in eq. \ref{pump} for both 
transition.  IR pumping rates only become comparable to $^{rot}A_{ul}$ for the 5--4 
transition if the total IR luminosity originates from a $\sim$2.5 pc source with a source 
temperature of $\gsim$100 K.   We conclude that IR pumping is not important 
for the 16--15 transition of HC$_{3}$N in any reasonable geometry of the IR field.  For 
the 5--4 transition to be sensitive to IR pumping, the IR source must be warm, opaque and 
extremely compact.  Therefore IR pumping is neglected for all clouds in the following discussion. 

\subsection{Physical Conditions of IC 342's Dense Gas --- LVG Modeling \label{lvgmod}}

The values of the excitation temperature constrain the
density and kinetic temperatures, $n_{H_{2}}$ and $T_k$, and the physical
conditions of the clouds driving the excitation. A series of Large
Velocity Gradient (LVG) radiative transfer models were run to predict
the observed intensities and line ratios for a given $n_{H_{2}}$,
T$_{k}$ and filling factor, $f_{a}=\Omega_{source}/\Omega_{beam}$, of
the dense component \citep[e.g.,][]{VLSW83}. Single component LVG
models are instructive, particularly when lines are optically thin, as
is the case for HC$_{3}$N. The LVG model used is that of
\citet{MTH00}, adapted to HC$_{3}$N, with levels up to J=20 included.
Collision coefficients are from \citet[][]{GC78}. A range of
densities, $n_{H_{2}}$ = 10$^{2}$--10$^{6}~cm^{-3}$, and kinetic
temperatures, T$_{k}$ = 0--100 K, was explored. HC$_{3}$N column
densities based on LTE excitation (Table \ref{Trat}), are calculated
at 2$^{''}$ resolution from: 

\begin{equation} 
N_{mol}~=~\left(\frac{3k Qe^{E_{u}/kT_{ex}}}{8\pi^{3}\nu S_{ul}\mu_{0}^{2}
g_{K_{u}}g_{I_{u}}} \right)I_{mol},
\end{equation}
using molecular data of \citet[][]{LL78}, HC$_{3}$N(5--4) intensities,
and T$_{ex}$(HC$_{3}$N) from Table \ref{Trat}. HC$_{3}$N abundances
are found to be, $X(HC_{3}N) \simeq 10^{-9.1}~-~10^{-8.5}$, with the
highest values towards C1 and D. While uncertain these abundances
agree with those found in \citet[][]{MT05} and are typical of Galactic
center HC$_{3}$N abundances \citep[e.g.][]{MTPZ76,DMNC00} and good
enough for constraining $X/dv/dr$.  The ratio of cloud linewidth to
core size is $\sim~1-3 \rm ~km~s^{-1}pc^{-1}$ for the GMCs.
Therefore, we adopt a standard model abundance per velocity gradient
of $X/dv/dr$ = 10$^{-9}$ km s$^{-1}$, but run models for values of
$X/dv/dr ~10^{-11}$ - $10^{-9}$.

Antenna temperatures are sensitive to the unknown filling factor.  In
these extragalactic observations, the beam corresponds to scales
large compared to cloud structure, and hence filling factors are not directly known.
To first order, the line ratios, $R_{10/5}$ and R$_{16/5}$
are independent of filling factor if we assume that HC$_{3}$N(5--4),
(10--9) and (16--15) originate in the same gas. So model ratios
are compared to the observed data to constrain parameter space.  For the 
parameter space implied by the line ratios model brightness temperatures 
are determined.  A comparison of the model brightness temperature to the 
observed brightness temperature sets the required areal filling factors for that 
solution.

\subsubsection{Physical Conditions of the Dense Component \label{dencomp}}

\begin{figure}
\epsscale{0.99}
\plotone{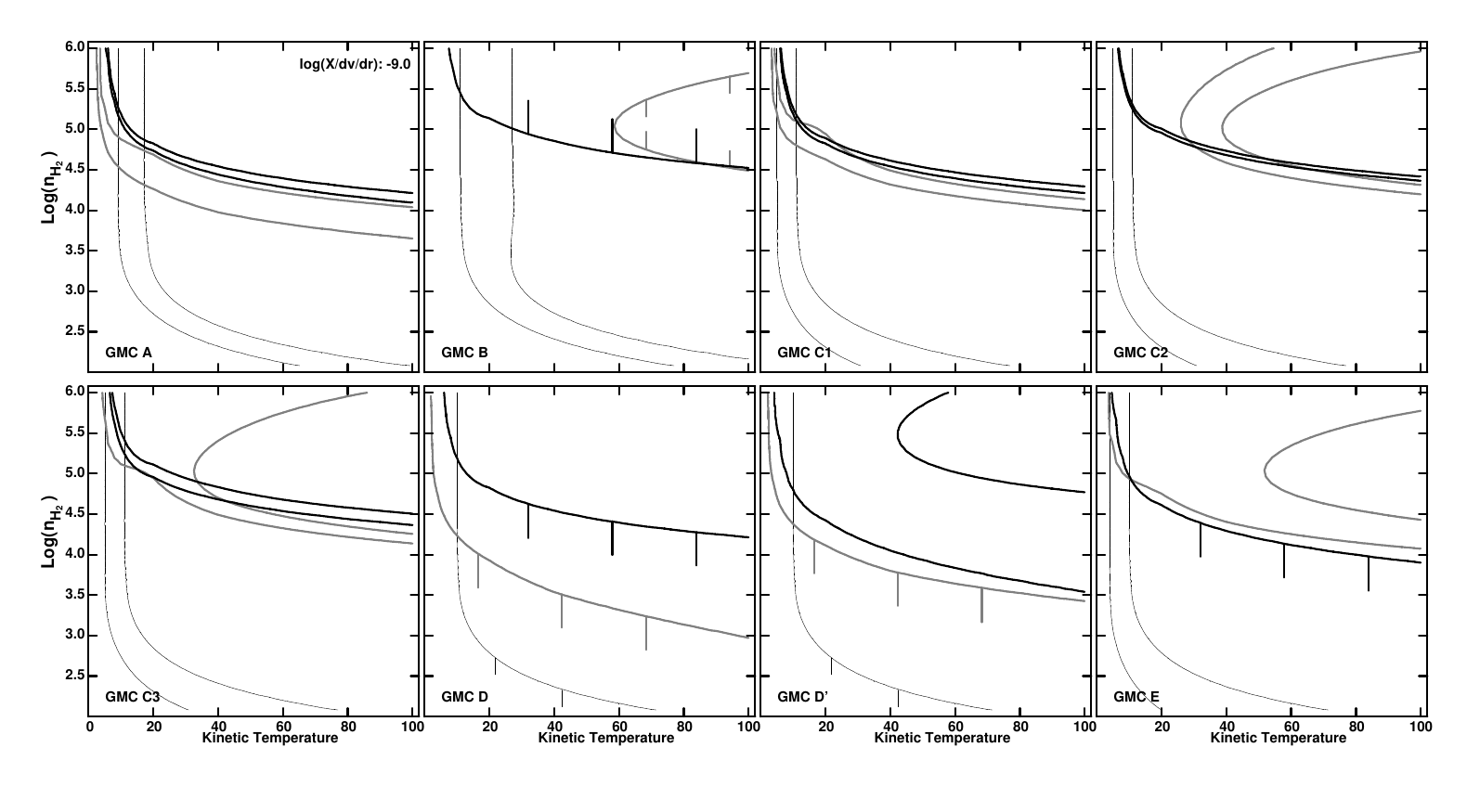}
\caption{{\it a)} The HC$_{3}$N LVG models of the GMCs in IC 342. The
thick black lines delineate the observed $\pm 1 \sigma$ range of
the HC$_{3}$N(16--15)/HC$_{3}$N(5--4) line ratio (R$_{16/5}$), sampled
at 2\arcsec\ resolution. The thick gray lines represents the observed
$\pm 1 \sigma$ range of the HC$_{3}$N(10--9)/HC$_{3}$N(5--4) line
ratio (R$_{10/5}$), sampled at 6\arcsec\ resolution.
$X(HC_{3}N)/dv/dr~=~10^{-9}$ is assumed for the displayed models. The 
thin black lines displays the $\pm 1 \sigma$ range implied by LVG models 
of C$^{18}$O for comparison \citep[][]{MT01}. Downward (upward) tickmarks 
indicate upper (lower) limits.
\label{lvg} }
\end{figure}

\begin{figure}
\epsscale{0.99}
\plotone{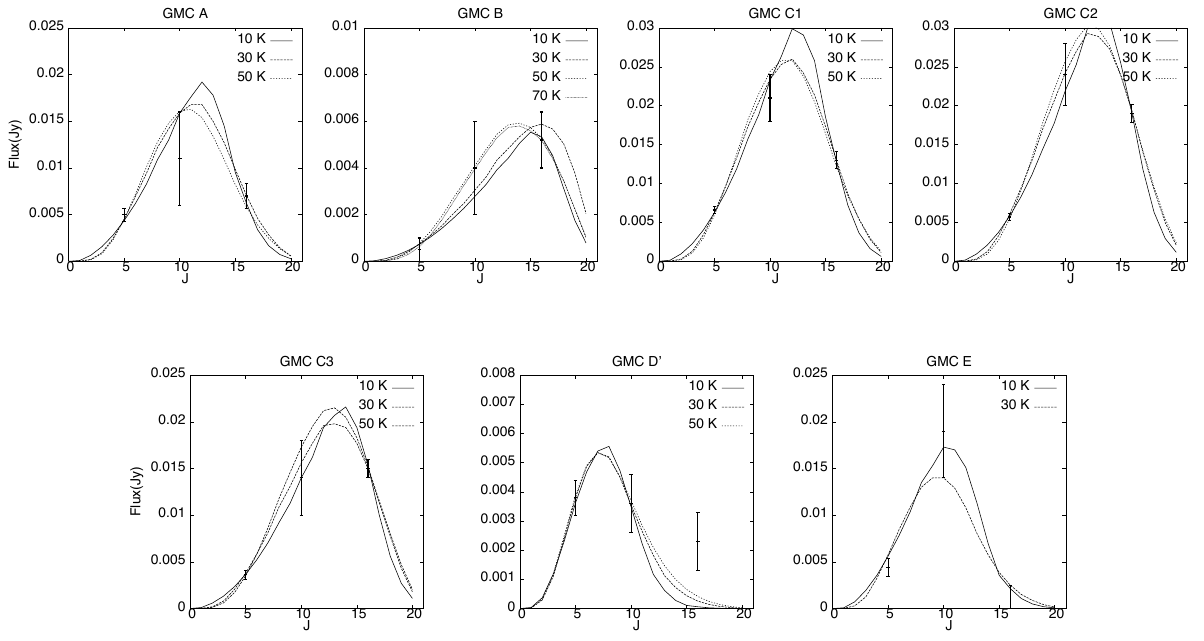}
\caption{HC$_{3}$N level population diagrams. Fluxes are taken from a
 2$^{''}$ box centered on the emission peak. HC$_{3}$N(10--9) fluxes
 are extrapolated from the (5--4) data using the line ratios in Table
 \ref{Tinti}. For each GMC, fitted LVG solutions are displayed for
 T$_K$ = 10 {\it (solid lines)}, 30 {\it (dashed lines)} and 50 K {\it
 (dotted lines)}, except GMC B which includes T$_{K}$ = 70 K {\it dot-dashed 
 lines)} (Table \ref{Tlvg1}). Only models with $X$(HC$_{3}$N)/$dv/dr$ = 10$^{-9}$
 km$^{-1}$ s are shown.  \label{levels} }
\end{figure}

Figure \ref{lvg} displays the results of the LVG modeling.  Acceptable 
$\pm1\sigma$ (T$_{k}$,$n_{H_{2}}$) parameter spaces are shown
for each line ratio (R$_{10/5}$, thick solid gray line; R$_{16/5}$, thick solid
black line). Also shown in Figure \ref{lvg} are the acceptable solutions 
obtained from the C$^{18}$O(2--1)/C$^{18}$O(1--0) line ratio \citep[thin gray lines;][]{MT01}. 
Figure \ref{levels} and Table \ref{Tlvg1} display the model flux versus
upper J state of the line for the T$_{K}$ = 10 K, 30 K and 50 K
solutions (also 70 K solutions for GMC B).

HC$_{3}$N(5--4) has a critical density of $n_{cr} \sim 5 \times
10^{4}$ cm$^{-3}$ and an upper energy state of 6.55 K.
HC$_{3}$N(16--15) has a critical density,  $n_{cr} \sim 5 \times10^{5}$ 
cm$^{-3}$ and an upper energy state of 59 K. HC$_{3}$N(10--9) values are
intermediate. Therefore $R_{10/5}$ and R$_{16/5}$ are sensitive probes
of $n_{H_{2}}$ between the range of $10^{4}$ and $10^{6}$
cm$^{-3}$. Observed ratios constrain $n_{H_{2}}$ to $\sim \pm$0.1 dex
at a given T$_{k}$ and to $\sim \pm$0.3 dex for all modeled T$_{k} >
20$ K. For T$_{k} ~\gsim$ 40 K, ratios are largely insensitive to
T$_{k}$ (curves are horizontal). The LVG models do not constrain density 
when T$_{k}$ is low. Thus to narrow the range of
possible solutions requires two or more line ratios, or an external
constraint on one of the axes. Kinetic temperatures of the dense
component at 2$^{''}$ scales in IC 342 remain largely unconstrained to
date. One method of constraining the kinetic temperature is the peak
T$_{b}$ at this resolution of the optically thick CO(2-1) line. Using
this method, T$_{k}~\simeq$ 35--45~K towards GMCs A, B and C, and
T$_{k}~\simeq$ 15--20 K for GMCs D, D\arcmin\ and E
\citep[][]{THH93,SBM03}. These values are not highly discordant from
arcminute resolution T$_{k}~\sim50$ K measurements from NH$_{3}$
\citep*[][]{HMR82} or the far-infrared dust temperature of 42 - 55 K 
\citep[][]{BGMNSWW80,RH84}. However, the CO(2-1) traces gas with 
densities two to three orders of magnitude lower than these HC$_{3}$N
transitions, so it is not clear that this is the relevant T$_{k}$ for
the dense cores of the GMCs.

Valid solutions are found for the GMCs, with the
exception of GMC D', which is the most uncertain due to its distance
from the center of the field. (No fit is attempted for GMC D because both
HC$_{3}$N(5--4) and HC$_{3}$N(16--15) are upper limits or tentative
detections.) The solutions show modest cloud excitations
given the strong nuclear star formation. Good agreement
is observed between solutions found using R$_{16/5}$ and R$_{10/5}$
when kinetic temperatures are low (T$_{k}~<30$ K). Given uncertainties
in line intensities and $X(HC_{3}N)/dv/dr$, higher temperatures cannot
be ruled out.  While T$_{k}$ itself is less well
constrained the combination $n_{H_{2}}$T$_{k}$ is well
determined. Since line brightnesses are the measured quantity, as gas
temperatures are raised, densities or filling factors must decrease to
compensate. The nature of the solutions are such that the densities
decrease more than the filling factors (Table \ref{lvg}).

GMCs A, C1 and C3 are best fit with $n_{H_{2}} \sim
10^{4.7-5.2}$~cm$^{-3}$ and T$_{k}~\simeq$ 20 K. For T$_{k} = 50$~K
derived densities drop to $n_{H_{2}} \sim 10^{4.3 - 4.7}$~cm$^{-3}$,
while they drop to $n_{H_{2}} \sim 10^{3.7 - 4.4}$~cm$^{-3}$ for
T$_{k}$ = 100 K. Changes in $X/dv/dr$ do not strongly influence the
derived solutions. Towards these clouds filling factors are
$f_{a}\simeq$0.03 - 0.1. For comparison, a 1 pc$^{2}$ cloud would
have $f_{a} ~\simeq$ 0.0012 for the aperture size and IC 342's
distance. Therefore these GMCs appear to have a few dozen dense
clumps similar to the larger clumps found in Galactic GMCs
\citep[e.g.][]{MB83,ZPT98}. GMC C2 appears to have similar densities
but slightly elevated temperatures, T$_{k}~\simeq$ 30K. In GMCs D and
D' densities are at least a factor of four lower than the other
clouds. Little can be said about the kinetic temperatures of these
GMCs. GMC D' is the only location with statistically discrepant
solutions from R$_{10/5}$ and R$_{16/5}$. Here the (16--15) line is
somewhat brighter than would be expected from the R$_{10/5}$. 

The starburst GMC, B, is markedly different from the others. The
combination of a low absolute HC$_{3}$N brightness temperature and
high ratios, $R_{10/5}$ and R$_{16/5}$, require higher densities of
$n_{H_{2}}\gsim 10^{5}$ cm$^{-3}$ and a smaller ($f_{a}<$0.02) filling
factor. Both line ratios seem to indicate a kinetic temperature of
more than T$_{k}~\simeq$ 40 K, and perhaps significantly higher,
averaged over our 30 pc beam. The dense gas in GMC B, which is
closest to the active star-forming region and IR source, is more
compact, denser, and more highly excited than the other clouds.

For our own Galactic Center it has long been argued that molecular
clouds must be unusually dense, at least $n\sim 10^4~\rm cm^{-3}$ to
withstand tidal forces \citep[eg.][]{stark89}. If one assumes that
GMC~C, the most well defined cloud, is a point mass in a spherical
mass potential of the galaxy, its Roche limit would be at a radius of
25~pc. The radius of the combined C1-C2-C3 complex is $>$25 pc. In
IC~342, the proximity to the nucleus where the strong noncircular
velocity field \citep{TH92} together with feedback from the starburst
\citep[][]{SBMC08} suggests that these clumps have densities large 
enough to maintain their identity but likely will not remain gravitationally 
bound to each other.  

\subsubsection{Comparisons Between HC$_{3}$N and C$^{18}$O Physical 
Conditions \label{compare}} 

For all GMCs (including GMC B),
T$_{ex}$(HC$_{3}$N) $\simeq$ T$_{ex}$(C$^{18}$O). If a model in which
all the H$_{2}$ exists in one uniform component, then the similarity
of T$_{ex}$ from both HC$_{3}$N and C$^{18}$O implies that the
densities of the molecular clouds are high enough ($>10^{4.5}$
cm$^{-3}$) to thermalize both C$^{18}$O and HC$_{3}$N across the
nucleus. (LVG solutions for C$^{18}$O J=2--1 and 1--0 lines
from \citet{MT01} are shown with the HC$_3$N solutions in Figure
\ref{lvg}.) In this monolithic model the low observed T$_{ex}$ of
10--20 K demand that the thermalized clouds must be quite cool. These
temperatures are similar to those of dark clouds in the disk of our
Galaxy, which is somewhat surprising given the elevated star formation
rate in the nucleus of IC~342.

However this model of a monolithic, dense, cold ISM seen in both
HC$_{3}$N and C$^{18}$O runs into problems with total mass
constraints. If the ISM is uniformly this dense and cold then the
total dense gas cloud mass implied by the required densities and
filling factors become very large, greater than permitted based on the
optically thin C$^{18}$O line emission. In Table
\ref{Tlvg1} the total mass of dense gas, $M_{den}$, is approximated from the 
LVG solutions assuming $M_{den}= m_{H_{2}}n_{H_{2}} R^{3}$, with R
defined as $R = \sqrt{f_{a}A_{beam}}$. Table \ref{Trat} lists the
total molecular gas mass, $M_{H_{2}}$, estimated from C$^{18}$O(2-1)
data \citep[see][]{MT01} assuming $M_{H_{2}} = m_{H_{2}}N_{H_{2}}
A_{beam}$. One can see that $M_{den}$ is typically larger than
$M_{H_{2}}$ for T$_{k}~<$ 30 K and diverges rapidly as T$_{k}$ drops
towards the LVG favored values. In short if GMCs have uniformly such
high densities and cold temperatures then the clouds would contain too
much H$_{2}$ mass for what is observed in C$^{18}$O. To match the 
line ratios while not violating mass constraints requires a multi-component 
dense ISM. A possible two component model of the dense gas is discussed 
in $\S$ \ref{cloudsxco}. 

\section{Dense Gas and the  CO Conversion Factor \label{xco}}

A consideration of the conversion factor between CO intensity and 
H$_{2}$ column density, $X_{CO}$, in the nuclear region of
IC 342 suggests that the model of a single, high density and
relatively cool ISM is not consistent with observations of CO 
isotopologues. Moreover these clouds are unlikely to resemble
Galactic disk GMCs in their internal structure and dynamics.

\subsection{The Single Component Model and $X_{CO}$ \label{mono}}

The well known Galactic
relation between CO intensity and H$_{2}$ column density,
$X_{CO}~=~I_{CO}/N(H_{2}),$ can be explained if GMCs consist of
optically thick (in CO) turbulent clumps in virial equilibrium
\citep[e.g.,][]{L81,SSBY87,SS87}. For the clumps to emit in HC$_{3}$N
densities must be large enough so that T$_{b}$ can legitimately be
approximated by T$_{k}$. If we adopt this model, then $X_{CO}$ is
approximated by:

\begin{equation}
X_{CO} \simeq 0.84\times 10^{20}  
 \left[\frac{\sqrt{\phi ~ ^{cl}n}}{f_{a} ~^{cl}T_{k}} \right]~\rm cm^{-2}\,
(K~km\,s^{-1})^{-1}
\label{eqxco}
\end{equation}
where $\phi$ is the volume filling factor and $^{cl}n$ is the density
of the {\it clumps} \citep[][]{S96,MB88}. From the brightness of the
CO isotopic lines, $^{CO}f_{a} \sim 1$ on 2\arcsec\ scales
\citep[][]{MT01}. The conversion factor $X_{CO}$ within the central
few hundred pc of IC 342 has been determined to be $0.6\times
10^{20}~\rm cm^{-2}\,(K~km\,s^{-1})^{-1}$
\citep[][]{MT01}. If the monolithic model presented in the previous section 
is correct and clouds are virialized clumps in equilibrium, then 
$\sqrt{^{cl}n}/T_{k}$ must be $ \lsim$0.75 to match
the observed conversion factor.

Table \ref{Tlvg1} records $\sqrt{^{cl}n}/T_{k}$ for each of the LVG
solutions across the nucleus of IC 342. For all solutions,
$\sqrt{^{cl}n}/T_{k}$ is greater than the required value, especially
at low T$_{k}$. For the single component LVG favored T$_{k}$ of $\sim$10 - 20 K towards
the more quiescent clouds, $\sqrt{^{cl}n}/T_{k} \sim $5--75, implying
$^{IC342}X_{CO}$ should be several times larger than $^{MW}X_{CO}$.  
However $X_{CO}$ in the nucleus of IC~342 is 3--4 times {\it lower} than the Galactic
conversion factor, based on both optically thin isotopologues of CO
as well as dust emission \citep{MT01}. Lower conversion factors appear
to be the norm for the nuclei of gas-rich star forming galaxies,
including our own \citep[eg.][]{SBMPN91,SYB97,DHWM98,HHR99,MT04,MTH08}. In
the nuclear region of IC 342 dense gas kinetic temperatures do not appear
to be higher than the Galactic disk value by an amount large enough to
offset the observed increase in density over typical Galactic
disk-like clouds; T$_{k}>$100~K would be required. Recently
\citet[][]{W07} completed a more generalized treatment of the
conversion factor including radiative transfer and concludes
that the true exponential dependences of $n_{H_{2}}$ and T$_{k}$ are
weaker than 0.5 and -1, respectively. However in all cases modelled,
the dependences remain positive for $n_{H_{2}}$ and negative for
T$_{k}$.

In summary, it is not possible to reconcile bright HC$_{3}$N
emission from uniformly cool, dense gas (low T$_{k}$, high $n_{H_{2}}$
and high $f_{a}$) with the known total amount of H$_{2}$ present if
virialized constant density clumps are adopted.

\subsection{A Possible Two Component Model and X$_{CO}$ \label{cloudsxco}}

The lack of an apparent connection between $\sqrt{^{cl}n}/T_{k}$
of the dense component and the observed conversion factor indicates
that the clouds in the nucleus of IC 342 cannot be treated as simple
virialized collections of uniform density clumps. One might expect
that a large fraction of the C$^{18}$O emission could originate from a
moderate density component that is distinct from the denser,
HC$_{3}$N-emitting gas. (Note that $^{12}$CO traces a distinct component 
more diffuse than that traced both by C$^{18}$O \citep[][]{MTH00} and HC$_{3}$N.)

As a first order extension to the basic virialized clumps model, we
imagine two spatially well-mixed sets of clumps; one low density
C$^{18}$O emitting, $n_{l}$, and one high density, $n_{h}$, that emits
both C$^{18}$O and HC$_{3}$N.  Assuming the fraction, by number, of
clumps with high density is $\Gamma$, then if both clumps have the
same T$_{k}$ the virialized clumps $X_{CO}$ relation becomes:

\begin{equation}
^{2comp}X_{CO} \simeq 0.84\times 10^{20}  
 \left[\frac{\sqrt{\phi ~ \tilde{n}}}{f_{a} ~T_{k}} \right]~\rm cm^{-2}\,
(K~km\,s^{-1})^{-1},
\label{eqxco2}
\end{equation}
where $\tilde{n} ~= ~ (1 - \Gamma)n_{l} + \Gamma n_{h}$.  The ratio of
dense gas to total H$_{2}$ becomes, M$_{den}$/M$_{tot}~\simeq \Gamma
n_{h}/((1-\Gamma)n_{l} + \Gamma n_{h})$ = $\Gamma n_{h}/\tilde{n}$.   For 
a given T$_{k}$, $n_{l}$ is chosen to match the
C$^{18}$O LVG solution \citep[][]{MT01}, while $n_{h}$ is chosen to
match the HC$_{3}$N LVG solutions, and $\Gamma~=~\phi_{HC3N}/\phi_{CO}
\simeq f_{a}^{3/2}$. Table \ref{Tlvg2} displays the adopted C$^{18}$O LVG solutions, along 
with the new $\sqrt{\tilde{n}}/T_{k}$ and M$_{den}$/M$_{tot}$ for the two component
model, assuming, for simplicity, that T$_{k}$ is the same for both
components.  $\sqrt{\tilde{n}}/T_{k}$ and hence $X_{CO}$ are
decreased from the simple one component model by factors of at least
four. This simple extension results in closer agreement between the
observed and calculated values of X$_{CO}$ (especially when including
the somewhat super-virial line widths), while maintaining most of the
mass in the dense (low filling factor) component.

It is almost certainly the case that this extension is an oversimplification.  
In reality, we expect the clouds to exhibit a continuum of densities.  However 
the LVG modeling demonstrates that at least these three components are 
required to match the multi-transition observations of $^{12}$CO, $^{13}$CO, 
C$^{18}$O and HC$_{3}$N.  This is consistent with conclusions from recent 
single-dish modeling of higher J transitions of HC$_{3}$N \citep[][]{AMMMB11}, 
but we find multiple components are required to match intensities not only between 
GMCs but within individual GMCs.

\section{HC$_{3}$N Emitting Dense Gas and Star Formation  \label{starf}}

\subsection{HC$_{3}$N versus HCN(1--0)  \label{starf_hcn}}

HCN(1--0) emission is the workhorse for relating quantities of dense gas to 
star formation \citep[e.g.][]{GS04}. It is interesting to compare conclusions about the dense 
component from HC$_{3}$N with those of the more commonly used HCN(1--0).  HCN(1--0) has been mapped 
at similar spatial resolution by \citet[][]{DRGGGM92}.  Figure \ref{inticont} compares HC$_{3}$N(5--4) 
to HCN(1--0).  While HC$_{3}$N generally 
traces the same dense GMCs seen in HCN their relative brightnesses are rather 
different.  In HCN(1--0) GMC A, B and C are all within 10 \% of the same brightness 
and GMCs D and E are 100 \% and 50 \% weaker, respectively.  Whereas in HC$_{3}$N(5--4) 
and (16--15), C dominates, B is nearly absent and A is not significantly different from 
D and E.  Comparisons with HC$_{3}$N clearly demonstrate that there is larger 
variations in dense gas properties than HCN indicates.  The dominate difference 
is enhanced HCN towards the starburst and GMC A relative to GMC C.  Unlike 
the HC$_{3}$N transitions, HCN(1--0) is optically thick and has slightly larger (25 - 50 \%) 
filling factors \citep[][]{DRGGGM92}.  As kinetic temperatures increase optically thick transitions 
brighten more rapidly than (lower excitation) optically thin transitions.  Hence it 
is expected that the HCN emission should favor somewhat warmer dense gas.  Likely this 
effect results in the much brighter relative HCN(1--0) intensities towards the starburst.  
The relative enhancement towards GMC A is less clear.  However, this cloud is known 
to be dominated by PDRs \citep[][]{MT05}, strongly influenced by mechanical feedback 
from the nuclear cluster \citep[][]{SBMC08} and has a complicated HC$_{3}$N temperature 
and velocity structure (Sections \ref{morp} and \ref{ratio}).  This cloud must have a complex 
density and temperature structure, potentially with a warmer intermediate density medium 
(Table \ref{Trat}).  Moreover, HCN abundances can be elevated in PDRs.
 
We conclude that while HCN(1--0) does a good job locating the dense gas but it does a poorer 
job tracing small scale variation in the properties of the dense gas when a mix of strong 
star formation and quiescent gas are present.  It is expected that such effects should become 
more important as specific star formation rates increase and spatial scales decrease.

\subsection{HC$_{3}$N versus Star Formation  \label{starf_rc}}

The brightest HC$_{3}$N emission and the brightest star-forming regions do not coincide.   
Most of the current star formation, traced by bright infrared 
and thermal radio continuum emission \citep{BGMNSWW80, TH83}, is situated about 50 
pc to the southwest of the dynamical center, in the vicinity of GMC B.   The brightest
HC$_{3}$N emission by far is on the northeast side of the nucleus, in the northern 
molecular arm centered at GMC C2. This region has free-free emission amounting to 
only a third of the brightness of the strongest radio source, associated with GMC B.  
The faintness, in absolute terms, of HC$_{3}$N(16--15) towards GMC B is unexpected and 
important reflecting a much lower areal filling of warm, very dense gas here.  Either the number 
of cloud clumps or their size is small relative to the other GMCs. 

On the other hand, the excitation of HC$_{3}$N does reflect the presence of young forming 
stars. HC$_{3}$N(16--15) and HC$_{3}$N(10--9) are relatively brighter towards the current 
starburst. This suggests that the absence of a correlation between HC$_{3}$N(5--4) intensity 
and star formation is partly due to depopulation of the lower energy transitions.  GMCs that are faint 
in the HC$_3$N J=16--15 transition and not associated with strong star formation show up well 
at the lower J transitions of HC$_{3}$N, as would be expected for gas of lower
excitation. Clearly there is a large amount of dense gas currently not actively forming 
stars, that shows up in the low excitation transitions of HC$_{3}$N. This is consistent with 
the fact that both HNC(1--0) and N$_{2}$H$^{+}$(1--0), generally considered dense quiescent 
gas tracers, are found to be very bright towards IC 342's nucleus \citep[][]{MT05}. 
HC$_{3}$N(5--4) appears to be an excellent extragalactic probe of the dense, quiescent 
molecular gas component not yet involved in the current starburst.

To quantitatively compare star formation with dense gas, a star formation rate is derived from the 
45 GHz continuum flux (spectral indices measured between 2 cm and 7 mm demonstrate that 
the vast majority of flux toward GMCs B and C is thermal Bremsstrahlung).  
The star formation rate is then compared with $M_{den}$ (Table \ref{Tlvg2}) 
to estimate a dense gas star formation efficiency, SFE$_{den} = SFR (M_{\odot}~yr^{-1})/M_{den}(M_{\odot})$.  
Dense gas depletion timescales, $\tau_{den}$= 1/SFE$_{den}$, are also computed.
In Table \ref{Tlvg2} the ratio of the observed Lyman continuum ionization rate 
\citep[e.g.][corrected for distance]{MT01,TTBCHM06} to M$_{den}$ derived from the LVG 
analysis is reported.  Towards the non-starburst GMCs $N_{Lyc}/M_{den} \simeq 1.6 - 4.0 \times 10^{45} 
s^{-1} M_{\odot}^{-1}$, while towards GMC B this ratio is 10 - 30 times larger.   If one adopts the 
conversion from N$_{Lyc}$ to star formation rate of $10^{-53}~ M_{\odot} ~yr^{-1} ~s^{-1}$ 
\citep[e.g.][]{K98}, then  SFE$_{den}$ for the starburst is $\sim 4 \times10^{-7} ~yr^{-1}$, or dense gas 
depletion times of $\tau_{den}\simeq$ 2-3 Myr! This highly enhanced SFE$_{den}$ is a direct 
consequence of the faint HC$_{3}$N emission here.  Even towards the non-starburst clouds 
SFE$_{den}$ are $1.6 - 4.0 \times 10^{-8}$ yr$^{-1}$.  These efficiencies are sufficiently short that they 
imply dense gas consumption timescales that are non-negligible fractions of the expected GMC lifetimes. 

The dense SFE is rather high across the nucleus, but the extreme value towards GMC 
B is remarkable.   \citet[][]{MT01} argued that intense star formation is suppressed along the 
spiral arms being triggered when the inflowing molecular gas collides with the inner 
ring molecular gas.  Therefore it is reasonable that in a relative sense SFE$_{den}$ 
is lower away from the central ring.  However this leaves unexplained why GMC B's SFE$_{den}$ is 
so much larger than GMC C, though its positions at the arm / inner ring intersection is the same. 
We suggest this is a sign of the evolution of star formation across the nucleus that is impacted by 
radiative and mechanical feed-back {\it from within} the molecular cloud.

\subsection{Destruction / Dispersal of Dense Gas with Starburst Age  \label{starf_dis}}
 
A possible cause of the different SFE$_{den}$ between GMC C and B is that we are 
observing the clouds at high enough spatial resolution to begin to identify the changing 
internal structure of the clouds in the presence of the starburst.  Over the lifetime 
of a cloud SFEs vary.  In the earliest stages of a star formation episode SFEs will appear 
low because elevated star formation rates have yet to convert the bulk of the molecular 
material to stars.  Towards the final stages of a GMCs evolution instantaneous SFEs appear 
to increase dramatically as the cloud clumps are consumed, destroyed or dispersed.  So 
observed instantaneous SFEs are expected to vary widely throughout the lifetime of an 
individual GMC and relative to lifetime averaged SFEs typically considered in extragalactic 
studies.  

If starburst B is a few Myrs more evolved than the other GMCs, especially 
the dynamical similar C, then we may be witnessing the consumption, 
dispersal or destruction phase of the remaining dense clumps in the presence 
of the expanding HII region.  The magnitude of the dense gas consumption 
times for B are indeed shorter than the lifetime of the GMC.  Several lines of 
evidence suggest that the (weaker) star formation towards C2 may 
be at a somewhat earlier phase.  These include less extended HII regions 
\cite[e.g.][]{TTBCHM06}, bright hot core-like species CH$_{3}$OH \citep[][]{MT05}, 
NH$_{3}$(6,6) \citep[][]{MHH06} and CO(6--5) \citep[e.g.][]{HHSGRG91}, and more 
mm dust continuum emission \citep[e.g.][]{MT01}.  

In this context it is interesting to compare the thermal pressure of the starburst HII 
region with that of the dense clumps along the same line of sight.  
Assuming a Str\"{o}mgren  sphere of R$\sim$3 pc, N$_{Lyc} = 1.1\times 10^{51}$ 
s$^{-1}$ and T$_{e}$ = 8000 K, parameters determined for the main starburst 
HII \citep[][]{TTBCHM06}, the thermal pressure of such an HII region would be 
$n_{e}$T$_{e}$ = $10^{6.76}$ cm$^{-3}$ K.  This value is equal given the 
uncertainties to $n_{H2}$T$_{K} \simeq 10^{6.57-6.75}$ cm$^{-3}$ K for the 
dense component of GMC B, hinting at pressure balance between the dense clumps 
and the HII region.  In addition, the filling factor of the HII region is larger than implied 
by the HC$_{3}$N LVG analysis.  So it is possible that HC$_{3}$N emission towards 
GMC B comes from dense clumps embedded within the HII region.

The above analysis suggests the following physical picture for the faint HC$_{3}$N 
emission towards the starburst.  The starburst towards B is more evolved.  The HII region 
at GMC B has had time to expand, destroying or dispersing the dense gas,
which is now in the form of smaller clumps and/or more diffuse gas.
Clumps that remain there must have high pressure to survive.
Near the younger star forming cloud GMC C (particularly toward GMC C2), 
the HII regions may just be developing, and have not had time to disperse
the clouds.  The dense clumps here would be more abundant and still present a 
hot core-like chemistry.  The clouds further from the central are on 
average less dense and at the current epoch remain largely quiescent, except 
possibly D', where the second high excitation component could be associated 
with the presence of shocks \citep[][]{MT05}.

\section{Conclusions \label{conc}}

We have imaged the HC$_3$N J=5--4 line in the nucleus of IC 342 with
the VLA and the HC$_{3}$N(16--15) line with the PdBI at $\lsim
2$\arcsec\ resolution. These are the first maps of these transitions
in an external galaxy. We have detected emission extended
along the nuclear ``mini-spiral" in (5--4) and more concentrated
emission in (16-15), with relative abundance of $X(HC_{3}N) ~\sim ~
1\times 10^{-9}$.   {\it HC$_{3}$N emission is not tightly correlated with
star formation strength.}  Dense gas {\it excitation} however, follows star
formation more closely. GMC B, which is weak in all the HC$_{3}$N lines, 
is relatively stronger in the higher J lines.

LVG modeling indicates that the HC$_3$N-emitting
gas has densities of 10$^{4-5}~\rm cm^{-3}$. In IC 342, physical
conditions of the densest component are fairly constant away from 
the immediate environment of the starburst, though beyond the central 
ring densities begin to fall.  Comparison with the C$^{18}$O observations of 
\citet{MT01} reveal excitation temperatures similar to C$^{18}$O values indicating
either that the molecular gas is dense and cool (T$_{k}~< 20$ K)
or that there are multiple gas components where the densities and
kinetic temperatures of each component conspire to give similar
overall excitation. The strong over-prediction of the amount of gas
mass present if densities are large and temperatures cool, favors a
multi-component ISM with at least two components beside the diffuse
one seen in $^{12}$CO(2--1). The actual ISM is likely a continuum of
cloud densities with different densities dominating different tracers.
HC$_{3}$N also differs in morphology from HCN(1--0), with HCN(1--0) 
being much brighter towards the starburst.  This is further evidence 
that there are multiple dense gas components.

Of particular interest, the starburst site (GMC B) exhibits the largest 
difference in intensity between HC$_{3}$N (both transitions) and HCN(1--0).  
The faintness of the HC$_{3}$N here suggests that the brightness of  
HCN(1--0) is not due solely to large quantities of dense gas.  A comparison 
between the GMCs with the largest star formation rates and similar 
dynamical environments, B and C, hint at an explanation.  While GMC B 
shows higher excitation, the low brightness of this cloud indicates that it is 
composed of a relatively small amount of warm, dense clumps.  
The smaller amount of dense gas at the site of the strongest young 
star formation indicate high star formation efficiencies in the dense gas. 
Towards GMC C, HC$_{3}$N, CH$_{3}$OH and NH$_{3}$ are more intense 
and the fraction of millimeter continuum from dust is higher.  This indicates 
that GMC C is in an early, less evolved (hot core-like) state.  The extreme 
dense gas star formation efficiency observed towards GMC B reflects the fact 
the main burst is in a more evolved state.  The dense clumps towards the 
starburst are being dispersed or destroyed in the presence of the HII region.  
The little dense gas remaining appears to be in pressure equilibrium with the 
HII region. The larger opacity of HCN(1--0) relative to HC$_{3}$N elevates its 
brightness temperature in this warm gas and lowers its critical density 
permitting it to remain excited in the somewhat lower density component.

We conclude that EVLA observations of HC$_{3}$N(5--4) can be a
powerful probe of dense, quiescent molecular gas in galaxies, and when
combined with high resolution imaging of the higher J transitions of
HC$_{3}$N with current and upcoming mm interferometers (like ALMA)
provide tight constraints on dense molecular gas properties in
stronger or more widespread starbursts, where changes like those
localized to GMC B are expected to permeate much of the ISM.

\acknowledgements 
 
We thank Chris Carilli and Miller Goss for assistance with the VLA
observations. We also thank Philippe Salome and Nemesio Rodriguez for
their help with the PdBI observations and data reduction.  We thank 
the referee for several suggestions that improved the presentation.
D.S.M. acknowledges support from the National Radio Astronomy
Observatory which is operated by Associated Universities, Inc., under
cooperative agreement with the National Science Foundation and NSF
grant AST-1009620.  This work is also supported by NSF grants
AST-0307950 and AST-0506469 to J.L.T.

\begin{deluxetable}{lcccccccc}
\tablenum{1}
\tablewidth{0pt}
\tablecaption{HC$_{3}$N  Measurements}
\tablehead{
\colhead{GMC}     & 
\colhead{I(5--4)\tablenotemark{a}}  &
\colhead{S$_{pk}$(5--4)} &
\colhead{$v_{o}$(5--4)} &
\colhead{$\Delta v$(5--4)} &
\colhead{I(16--15)\tablenotemark{a}} &
\colhead{S$_{pk}$(16--15)} &
\colhead{$v_{o}$(16--15)} &
\colhead{$\Delta v$(16--15)}  \\
\colhead{} &
\colhead{\footnotesize($\rm K~ km~ s^{-1}$)} &
\colhead{\footnotesize($\rm mJy$)} &
\colhead{\footnotesize($\rm km~s^{-1}$)} &
\colhead{\footnotesize($\rm km~s^{-1}$)} &
\colhead{\footnotesize($\rm K~ km~ s^{-1}$)} &
\colhead{\footnotesize($\rm mJy$)} &
\colhead{\footnotesize($\rm km~s^{-1}$)} &
\colhead{\footnotesize($\rm km~s^{-1}$)} 
}
\startdata
A &13$\pm$2.6  &5.0$\pm$0.7  &22$\pm$2 &29$\pm$5 &2.4$\pm$0.5 &
 7.0$\pm$1.3 & 16$\pm$3 & 44$\pm$10 \\
B &$<$5.3  &$<$1.0  &\nodata&\nodata &2.4$\pm$0.5 &
 5.2$\pm$1.5 & 19$\pm$3 & 28$\pm$10 \\
C1 &44$\pm$2.6  &6.6$\pm$0.5  &48$\pm$1 &39$\pm$3 &8.5$\pm$0.5 &
 13$\pm$1.1 & 47$\pm$1 & 43$\pm$4 \\
C2 &28$\pm$2.6  &5.7$\pm$0.5  &50$\pm$1 &35$\pm$4 &10$\pm$0.5 &
 19$\pm$1.2 & 45$\pm$1 & 47$\pm$4 \\
C3 &14$\pm$2.6  &3.6$\pm$0.5  &49$\pm$1 &31$\pm$5 &2.8$\pm$0.5 &
 15$\pm$1.0 & 45$\pm$1 & 47$\pm$4 \\
D &13$\pm$2.6  &1.5$\pm$0.5  &51$\pm$6 &58$\pm$20 &$<$1.0 &
 $<$2.5 & \nodata & \nodata \\
D' &20$\pm$2.6  &3.8$\pm$0.5  &52$\pm$2 &40$\pm$7 &$\lsim$1.0 &
 2.3$\pm$1.1 & 57$\pm$8 & 44$\pm$26 \\
E &14$\pm$2.6  &4.4$\pm$1  &12$\pm$2 &28$\pm$7 &$<$1.0 &
 $<$2.5 &\nodata & \nodata \\   
 
\enddata
\tablecomments{Based on spectra from 2$^{''}$ apertures
centered on each cloud, except where noted.  Uncertainties in the
temperatures and intensities are the larger of the rms or 10 \%
absolute calibration uncertainties.  Uncertainties for the line
centroids and widths are the 1$\sigma$ uncertainties in the gaussian
fits.}
\tablenotetext{a}{Based on the full resolution  data.}
\label{Tinti}
\end{deluxetable}

\begin{deluxetable}{lcccccccc}
\tablenum{2}
\tablewidth{0pt}
\tablecaption{Line Ratios and Excitation}
\tablehead{
\colhead{GMC}     & 
\colhead{$R_{10/5}$\tablenotemark{a}}  &
\colhead{$^{10/5}$T$_{ex}$\tablenotemark{a}} &
\colhead{$R_{16/5}$\tablenotemark{b}}  &
\colhead{$^{16/5}$T$_{ex}$\tablenotemark{b}} &
\colhead{$^{C18O}$T$_{ex}$\tablenotemark{c}} &
\colhead{N(H$_{2}$)\tablenotemark{c}} &
\colhead{X(HC$_{3}$N)} &
\colhead{$^{C18O}$M$_{H2}$\tablenotemark{c}}  \\
\colhead{} &
\colhead{}  &
\colhead{(K)} &
\colhead{}  &
\colhead{(K)} &
\colhead{(K)} &
\colhead{(cm$^{-2}$)} &
\colhead{} &
\colhead{($M_{\odot}$)} 
}
\startdata
A &0.55$\pm$0.2  &8.8$\pm$1.5  &0.14$\pm$0.03 &12$\pm$0.6 &13$\pm$4 &
 4(22) & 1(-9) & 4.4(5) \\
B &2.5$\pm$1  &37$\pm$19  &$\gsim$0.52 & $\gsim$18 &19$\pm$8 &
 4(22) & $>$8(-10) & 7.6(5) \\
C1 &0.80$\pm$0.1  &11$\pm$1.0  &0.20$\pm$0.03 &13$\pm$0.6 &8$\pm$3\tablenotemark{d} &
 7(22) & 3(-9) & 6.8(5) \\
C2 &1.1$\pm$0.1  &14$\pm$1.0  &0.33$\pm$0.03 &15$\pm$0.6 &8$\pm$3 &
 1(23) & 2(-9) & 1.0(6) \\
C3 &1.0$\pm$0.1  &13$\pm$1.3  &0.42$\pm$0.07 &17$\pm$1.3 &8$\pm$3\tablenotemark{d} &
 5(22) & 1(-9) &4.9(5)  \\
D &$\lsim$0.15  &$\sim$5  & $<$0.17 &$<$13 &6$\pm$4 &
 2(22) & 3(-9) & 6.4(5) \\
D' &$\sim$0.24  &$\sim$6  &0.061$\pm$0.04 &10$\pm$1.8 &$\sim$6 &
 6(22) & 1(-9) & 5.8(5)  \\
 E &1.1$\pm$0.3  &14$\pm$3.1  &$<$0.057 &$<$10 &7$\pm$3 &
 6(22) & 1(-9) &1.1(6) \\
\enddata
\tablecomments{Based on spectra from 2$^{''}$ apertures
centered on each cloud, except where noted.  Uncertainties in the
temperatures and intensities are the larger of the rms or 10 \%
absolute calibration uncertainties.  Uncertainties for the line
centroids and widths are the 1$\sigma$ uncertainties in the gaussian
fits.}
\tablenotetext{a}{Based on 6\arcsec\ data.}
\tablenotetext{b}{Based on the full resolution data.}
\tablenotetext{c}{From \citet[][]{MT01}.}
\tablenotetext{d}{Assumed to be constant across GMC C.}
\label{Trat}
\end{deluxetable}

\begin{deluxetable}{lccccccc}
\tablenum{3}
\tablewidth{0pt}
\tablecaption{One Component HC$_{3}$N LVG Solutions ($X/dv/dr = 10^{-9}$)}
\tablehead{
\colhead{GMC}     & 
\colhead{T$_{k}$}  &
\colhead{log($n_{H_{2}}$)} &
\colhead{$f_{a}$} &
\colhead{$\frac{\sqrt{n_{H_{2}}}}{T_{k}}$} &  
\colhead{$M_{den}$\tablenotemark{a}} &
\colhead{$\frac{M_{den}}{M_{H2}}$} &
\colhead{log(n$_{H2}$T$_{k}$)}\\
\colhead{} &
\colhead{($\rm K$)} &
\colhead{($\rm log (cm^{-3}$))} &
\colhead{} &
\colhead{} &
\colhead{$\rm (M_{\odot}$)} &
\colhead{} &
\colhead{(log($cm^{-3}K$))} 
}
\startdata
 A &10 & 5.15 & 0.091 & 38  & 4.7(6)\tablenotemark{b} & 11& 6.16 \\
    & 30 & 4.63 & 0.051 & 6.9 & 6.0(5) &1.4 & 6.11 \\
    & 50 & 4.39 & 0.067 & 3.1 & 5.2(5) &1.2 & 6.09\\
B &10 & 5.75 & 0.015 &  75  & 1.3(6) &1.7 & 6.75 \\
    & 30 & 5.22 & 4.7(-3) & 14 & 6.5(4) &0.084 & 6.70 \\
    & 50 & 4.87 & 6.0(-3) & 5.5 & 4.2(4) & 0.055 & 6.57 \\
    & 70 & 4.81 & 5.4(-3) & 3.6 & 3.1(4) & 0.041 & 6.66 \\
C1 &10 & 5.23 & 0.13 & 41  & 9.7(6) &14 & 6.23 \\
    & 30 & 4.69 & 0.066 &7.4 & 1.0(6) &1.5 & 6.17 \\
    & 50 & 4.50 & 0.075 & 3.6 & 8.0(5) &1.2  &6.19  \\
C2 &10 & 5.36 & 0.12 & 48  & 1.3(7) &13 & 6.36 \\
    & 30 & 4.80 & 0.056 & 8.4 & 1.0(6) & 1.0& 6.28 \\
    & 50 & 4.64 & 0.057 & 4.2 & 7.3(5) &0.73 & 6.34 \\
C3 &10 & 5.44 & 0.076 &53  & 7.1(6) &14 & 6.44 \\
    & 30 & 4.86 & 0.033 & 9.0 & 5.3(5) &1.1 & 6.34 \\
    & 50 & 4.70 & 0.034 & 4.5 & 3.8(5) &0.0.76 & 6.40 \\
D &10 & $<$4.49 & $>$0.040 &  \nodata  & \nodata &  \nodata  & $<$5.49  \\
    & 30 & $<$4.10 & $>$0.040 & \nodata & \nodata &  \nodata  & $<$5.58 \\
    & 50 & $<$3.88 & $>$0.060  & \nodata & \nodata  &  \nodata  & $<$5.58 \\
D' &10 & \nodata & \nodata & \nodata & \nodata & \nodata & \nodata \\
    & 30 & \nodata & \nodata & \nodata & \nodata & \nodata & \nodata \\
    & 50 & \nodata & \nodata & \nodata & \nodata & \nodata & \nodata \\
E &10 & 4.90 & 0.12 & 28  & 4.0(6) &3.6 & 5.90 \\
    & 30 & 4.36 & 0.096 & 5.1 & 8.3(5) &0.75 & 5.84\\
    & 50 & \nodata & \nodata & \nodata & \nodata & \nodata & \nodata \\
\enddata
\tablenotetext{a}{$M_{dense}$ is an estimate of the dense H$_{2}$ mass implied 
by these LVG solutions.  $M_{den} ~=~m_{H_{2}}n_{H_{2}}R_{e}^{3}$, with $R_{e}$ 
defined as $R_{e}^{2} ~=~ f_{a}A_{aper}$. }
\tablenotetext{b}{Numbers of the form a(b) are equal to $a \times 10^{b}$.}
\label{Tlvg1}
\end{deluxetable}

\clearpage 

\begin{deluxetable}{lccccccc}
\tablenum{4}
\tablewidth{0pt}
\tablecaption{Two Component HC$_{3}$N LVG Solutions ($X/dv/dr = 10^{-9}$)}
\tablehead{
\colhead{GMC}     & 
\colhead{$^{both}$T$_{k}$}  &
\colhead{log($n_{\ell}$)} &
\colhead{$\Gamma$} &
\colhead{$log(\tilde{n})$} &  
\colhead{$\frac{\sqrt{\tilde{n}}}{T_{k}}$} &  
\colhead{$\frac{M_{den}}{M_{H2}}$}&
\colhead{$\frac{N_{Lyc}}{M_{den}}$} \\
\colhead{} &
\colhead{($\rm K$)} &
\colhead{($\rm log (cm^{-3}$))} &
\colhead{} &
\colhead{($\rm log (cm^{-3}$))} &
\colhead{} &  
\colhead{} &
\colhead{($\rm s^{-1} M_{\odot}^{-1}$)} 
}
\startdata
 A & 30 & 2.6 & 0.015 & 3.01 & 1.1 &0.62  &4.0(45)\\
    & 50 & 2.4 & 0.017 & 2.82 & 0.55 &0.63 &4.0(45) \\
B  & 30 &2.8 & 3.2(-4) & 2.83 & 0.87 &0.078 & 5.2(46)\\
    & 50 & 2.5 & 4.6(-4) & 2.54 & 0.37 & 0.097 & 4.2(46)  \\
    & 70 & 2.4 & 4.0(-4) & 2.44 & 0.24 & 0.092 & 4.4(46)\\
C1& 30 & 2.4 & 0.017 &3.03 & 1.1 &0.77 &1.9(45) \\
    & 50 & 2.1 & 0.021 & 2.90 & 0.56 &0.84 &1.8(45) \\
C2& 30 & 2.4 & 0.021 & 3.03 & 1.1 & 0.77 &2.1(45)\\
    & 50 & 2.1 & 0.013 & 2.87 & 0.54 &0.83 &1.9(45)\\
C3& 30 & 2.3 & 5.0(-3) & 2.80 & 0.84 &0.69 &4.1(45)\\
    & 50 & 2.1 & 6.3(-3) & 2.64 & 0.42 &0.72 &4.0(45)\\
D & 30 & $<$2.5 &\nodata & \nodata & \nodata &  \nodata &\nodata  \\
    & 50 & $<$2.3 &\nodata  & \nodata & \nodata  &  \nodata &\nodata  \\
D' & 30 & \nodata & \nodata & \nodata & \nodata & \nodata &\nodata \\
    & 50 & \nodata & \nodata & \nodata & \nodata & \nodata &\nodata \\
E & 30 & 2.3 & 0.030 & 2.94 & 0.99 &0.78 & 1.6(45) \\
    & 50 & 2.0 & \nodata & \nodata & \nodata & \nodata & \nodata  \\
\enddata
\tablecomments{The low density component comes from the LVG solutions to C$^{18}$O 
\citep[][]{MT01}.  For sake of simplicity it is assumed that the kinetic temperature of both 
the low and high density components are the same.  See section \ref{cloudsxco} for discussion 
of $\Gamma$ and $\tilde{n}$.}
\label{Tlvg2}
\end{deluxetable}

\begin{thebibliography}{} 

\bibitem[Aalto et al.(2002)]{APHC02} Aalto, S., Polatidis, A.~G., H{\"u}ttemeister, S., 
\& Curran, S.~J.\ 2002, \aap, 381, 783 
\bibitem[Aladro et al.(2011)]{AMMMB11} Aladro, R., Mart{\'{\i}}n-Pintado, J., Mart{\'{\i}}n, S., 
Mauersberger, R., \& Bayet, E.\ 2011, \aap, 525, A89 
\bibitem[Becklin et al.(1980)]{BGMNSWW80} Becklin, E. E., Gatley, I., 
Mathews, K., Neugebauer, G., Sellgren, K., Werner, M. K., \& Wynn-Williams, 
C. G.1980, \apj, 236, 441
\bibitem[Brandl et al.(2006)]{Brandl06}Brandl, B. R., et al. 2006, \apj, 653, 1129
\bibitem[Bussmann et al.(2008)]{Bet08} Bussmann, R.~S., et 
al.\ 2008, \apjl, 681, L73 
\bibitem[Costagliola \& Aalto(2010)]{CA10}Costagliola, F. \& Aalto, S., 2010, \aap, 515, 71
\bibitem[Dahmen et al.(1998)]{DHWM98} Dahmen, G., 
Huttemeister, S., Wilson, T. L., \& Mauersberger, R. 1998, \aap, 331, 959 
\bibitem[de Vicente et al.(2000)]{DMNC00} de Vicente, P., Mart{\'{\i}}n-Pintado, J., Neri, R., \& 
Colom, P.\ 2000, \aap, 361, 1058 
 \bibitem[Downes et al.(1992)]{DRGGGM92} Downes, D., Radford, S. J. E., 
Giulloteau, S., Guelin, M., Greve, A., \& Morris, D. 1992, \aap, 262, 424
\bibitem[Fuente et 
al.(1993)]{FMCB93} Fuente, A., Martin-Pintado, J., Cernicharo, J., \& 
Bachiller, R.\ 1993, \aap, 276, 473 
 \bibitem[Gao \& Solomon(2004)]{GS04}Gao, Y. \& Solomon, P. M. 2004,  \apj,
 606, 271
 \bibitem[Garc{\'i}a-Burillo et al.(2000)]{GMFN00}Garc{\'i}a-Burillo, S., 
Mart{\'i}n-Pintado, J.,  Fuente, A. \&  Neri, R. 2000, \aap, 355, 499
 \bibitem[Garc{\'i}a-Burillo et al.(2001)]{GMFN01}Garc{\'i}a-Burillo, S., 
Mart{\'i}n-Pintado, J.,  Fuente, A. \&  Neri, R. 2001, \apj, 563, L27
 \bibitem[Garc{\'i}a-Burillo et al.(2002)]{GMFUN02}Garc{\'i}a-Burillo, S., 
Mart{\'i}n-Pintado, J.,  Fuente, A., Usero, A. \&  Neri, R. 2002, \apj, 575, L55
\bibitem[Graci{\'a}-Carpio et al.(2008)]{GGPFU08} Graci{\'a}-Carpio, J., 
Garc{\'{\i}}a-Burillo, S., Planesas, P., Fuente, A., \& Usero, A.\ 2008, \aap, 479, 703 
 \bibitem[Green \& Chapman(1978)]{GC78}Green, S. \& Chapman, S. 1978, \apjs, 
 37, 169
 \bibitem[Harris et al.(1991)]{HHSGRG91}Harris, A. I., Hills, R. E., Stutzki, J., Graf, U. U., 
 Russell, A. G. \& Genzel, R., 1991, \apjl, 382, L75
\bibitem[Harrison et al.(1999)]{HHR99} Harrison, A., Henkel, 
C., \& Russell, A. 1999, \mnras, 303, 157
   \bibitem[Ho et al.(1982)]{HMR82} Ho, P. T. P., Martin, R. N., 
\& Ruf, K. 1982, \aap, 113, 155 
  \bibitem[Hunter et al.(1997)]{Het97}Hunter, S. D. et al. 1997, \apj, 
481, 205 
 \bibitem[Ishizuki et al.(1990)]{I90} Ishizuki, S., Kawabe, R., Ishiguro, 
M., Okumura, S. K., Morita, K.-I., Chikada, Y., \& Kasuga, T. 1990, \nat, 
344, 224
\bibitem[Kennicutt(1998)]{K98} Kennicutt, R.~C., Jr.\ 1998, \araa, 36, 189 
\bibitem[Knudsen et al.(2007)]{KWWBRM07} Knudsen, K.~K., Walter, 
F., Weiss, A., Bolatto, A., Riechers, D.~A., \& Menten, K.\ 2007, \apj, 666, 156 
\bibitem[Krips et al.(2008)]{KNGMCGE08} Krips, M., Neri, R., 
Garc{\'{\i}}a-Burillo, S., Mart{\'{\i}}n, S., Combes, F., 
Graci{\'a}-Carpio, J., \& Eckart, A.\ 2008, \apj, 677, 262 
   \bibitem[Karachentsev(2005)]{K05} Karachentsev, I. D. 2005, \aj, 129, 178 
 \bibitem[Lafferty \& Lovas(1978)]{LL78}Lafferty, W. J. \& Lovas, F. J. 1978, 
 J. Phys. Chem. Ref. Data, 7, 441
\bibitem[Larson(1981)]{L81} Larson, R. B. 1981, \mnras, 194, 809
 \bibitem[Lo et al.(1984)]{Letal84}Lo, K. Y. et al. 1984, \apj, 282, L59
  \bibitem[Maloney \& Black(1988)]{MB88} Maloney, P., \& Black, J. H. 
1988, \apj, 325, 389
 \bibitem[Mauersberger et al.(1990)]{MHS90}Mauersberger, R., Henkel, 
 C. \& Sage, L. J. 1990, \aap, 236, 63
  \bibitem[Meier \& Turner(2001)]{MT01}Meier, D. S. \& Turner, J. L. 
2001, \apj, 551, 687
  \bibitem[Meier \& Turner(2004)]{MT04} Meier, D. S., \& Turner, J. L.  2004, 
\aj, 127, 2069 
  \bibitem[Meier \& Turner(2005)]{MT05}Meier, D. S. \& Turner, J. L. 
2005, \apj, 618, 259
  \bibitem[Meier et al.(2000)]{MTH00}Meier, D. S., Turner, J. L. 
\& Hurt, R. L. 2000, \apj, 531, 200
\bibitem[Meier et al.(2008)]{MTH08} Meier, D.~S., Turner, J.~L., \& Hurt, R.~L.\ 2008, 
\apj, 675, 281 
\bibitem[Montero-Casta{\~n}o et al.(2006)]{MHH06} Montero-Casta{\~n}o, M., Herrnstein, R.~M., 
\& Ho, P.~T.~P.\ 2006, \apj, 646, 919 
  \bibitem[Morris et al.(1976)]{MTPZ76}Morris, M., Turner, B. E., Palmer, P. 
  \& Zuckerman, B. 1976, \apj, 205, 82
\bibitem[Myers \& Benson(1983)]{MB83} Myers, P.~C., \& Benson, P.~J.\ 1983, \apj, 266, 309 
\bibitem[Narayanan et al.(2008)]{NCSDHW08} Narayanan, D., Cox, 
T.~J., Shirley, Y., Dav{\'e}, R., Hernquist, L., 
\& Walker, C.~K.\ 2008, \apj, 684, 996 
\bibitem[Papadopoulos(2007)]{P07} Papadopoulos, P.~P.\ 
2007, \apj, 656, 792 
   \bibitem[Rickard \& Harvey(1984)]{RH84} Rickard, L. J, \& Harvey, P. 
M. 1984, \aj, 89, 1520
 \bibitem[Saha et al.(2002)]{SCH02}Saha, A., Claver, J. \& 
      Hoessel, J. G. 2002, \aj, 124, 839
  \bibitem[Sakamoto(1996)]{S96} Sakamoto, S. 1996, \apj, 462, 215
  \bibitem[Schinnerer et al.(2003) ]{SBM03}Schinnerer,
    E., B\"{o}ker, T., \& Meier, D. S. 2003, \apjl, 591, L115
    \bibitem[Schinnerer et al.(2008)]{SBMC08} Schinnerer, E., 
B{\"o}ker, T., Meier, D.~S., \& Calzetti, D.\ 2008, \apjl, 684, L21 
\bibitem[Schulz et al.(2001)]{SGKK01}Schulz, A., G\"{u}sten, R.,
  K\"{o}ster, B \& Krause, D. 2001, \aap, 371, 25  
\bibitem[Scoville \& Sanders (1987)]{SS87} Scoville, N. Z., \&
  Sanders, D. B. 1987, in Interstellar Processes, ed. D. J. Hollenbach
and H. A. Thronson, Jr., (Dordrecht: Reidel), 21
  \bibitem[Scoville et al.(1997)]{SYB97} Scoville, N. Z., Yun, 
M. S., \& Bryant, P. M. 1997, \apj, 484, 702 
\bibitem[Smith et al.(1991)]{SBMPN91} Smith, P. A., Brand, 
P. W. J. L., Mountain, C. M., Puxley, P. J., \& Nakai, N. 1991, \mnras, 
252, 6
  \bibitem[Solomon et al.(1987)]{SSBY87}
Solomon, P. M., Rivolo, A. R., Barrett, J., \& Yahil, A. 1987, \apj, 319, 730
  \bibitem[Stark et al. (1989)]{stark89}
Stark, A. A., Bally, J., Wilson, R. W., \& Pound, M. W. 1989, in The Center of
the Galaxy, ed. M. Morris, (Dordrecht: Kluwer), 129
  \bibitem[Strong et al.(1988)]{Set88} Strong, A. W. et al., 1988, 
\aap, 207, 1
  \bibitem[Tsai et al.(2006)]{TTBCHM06}Tsai, C.-W., Turner, J. L.,
 Beck, S. C., Crosthwaite, L. P., Ho, P. T. P. \& Meier, D. S. 2006, 
\aj, in press
  \bibitem[Turner \& Ho(1983)]{TH83} Turner, J. L., \& Ho, P. T. P. 1983, 
\apj, 268, L79
  \bibitem[Turner \& Hurt(1992)]{TH92} Turner, J. L., Hurt,
    R. L. 1992, \apj, 384, 72
  \bibitem[Turner et al.(1993)]{THH93} Turner, J. L., Hurt, 
R. L., \& Hudson, D. Y. 1993, \apj, 413, L19
    \bibitem[Usero et al.(2004)]{UGFMR04}Usero, A., Garc{\'i}a-Burillo,
      S., Fuente, A., Mart{\'i}n-Pintado, J. \& Rodr{\'i}guez-Fern{\'a}ndez,
      N. J. 2004, \aap, 419, 897
  \bibitem[Usero et al.(2006)]{UGMFN06} Usero, A., Garc{\'{\i}}a-Burillo, 
S., Mart{\'{\i}}n-Pintado, J., Fuente, A., \& Neri, R. 2006, \aap, 448, 457 
 \bibitem[Vanden Bout et al.(1983)]{VLSW83}Vanden Bout, P.  A., Loren, 
R. B., Snell,  R. L. \& Wootten, A. 1983, \apj, 271, 161
\bibitem[Wall(2007)]{W07} Wall, W.~F.\ 2007, \mnras, 379, 
674 
\bibitem[Wu et al.(2005)]{WEGSSV05} Wu, J., Evans, N.~J., II, 
Gao, Y., Solomon, P.~M., Shirley, Y.~L., 
\& Vanden Bout, P.~A.\ 2005, \apjl, 635, L173 
\bibitem[Zinchenko et al.(1998)]{ZPT98} Zinchenko, I., Pirogov, L., \& 
Toriseva, M.\ 1998, \aaps, 133, 337 



\end{thebibliography}
\end{document}